\documentstyle[12pt]{article}

\begin{document}

\newcommand{\beq}{\begin{equation}}
\newcommand{\eeq}{\end{equation}}
\newcommand{\barr}{\begin{eqnarray}}
\newcommand{\earr}{\end{eqnarray}}

\newcommand{\andy}[1]{ }

\def\Da{D_\alpha}
\def\rz{\mbox{\boldmath $y$}_0}
\def\hh{\widehat}
\def\tt{\widetilde}
\def\cH{{\cal H}}
\def\l{\langle}
\def\r{\rangle}
\def\txt{\textstyle}
\def\bmom{\mbox{\boldmath $\omega$}}
\def\bmR{\mbox{\boldmath $R$}}
\def\bmbeup{\mbox{\boldmath $^\beta$}}
\def\bmbedown{\mbox{\boldmath $_\beta$}}
\def\bmsigup{\mbox{\boldmath $^\widehat{\Sigma}$}}
\def\bmprod{\mbox{\boldmath $^{\beta\cdot\widehat{\Sigma}}$}}


\def\ask{\marginpar{?? ask:  \hfill}}
\def\fin{\marginpar{fill in ... \hfill}}
\def\note{\marginpar{note \hfill}}
\def\check{\marginpar{check \hfill}}
\def\discuss{\marginpar{discuss \hfill}}

\begin{titlepage}
\begin{flushright}
\today \\
BA-TH/96-246\\
WU-HEP-96-9\\
\end{flushright}
\vspace{.5cm}
\begin{center}
{\LARGE Emergence of a Wiener process as a result of the quantum
  mechanical interaction with a macroscopic medium }\\ [.5cm]

{\large Raffaella BLASI,$^{(1)}$ Hiromichi NAKAZATO,$^{(2)}$ \\
     Mikio NAMIKI,$^{(2)}$ and Saverio PASCAZIO$^{(1,3)}$ \\
           \quad    \\
        $^{(1)}$Dipartimento di Fisica, Universit\`a di Bari \\
I-70126  Bari, Italy \\
        $^{(2)}$Department of Physics,
Waseda University \\ Tokyo 169, Japan \\
        $^{(3)}$Istituto Nazionale di Fisica Nucleare, Sezione di Bari \\
 I-70126  Bari, Italy

}
\end{center}
\vspace*{.5cm}
\vspace*{.5cm}
\vspace*{.5cm}

\begin{center}
{\small\bf Abstract}\\ \end{center}

{\small
We analyze a modified version of the Coleman-Hepp model, that is able to take
into account energy-exchange processes between the incoming particle
and the linear array made up of $N$ spin-1/2 systems. We bring to light the
presence of a Wiener dissipative process in the weak-coupling, macroscopic ($N
\rightarrow \infty$) limit. 
In such a limit and in a restricted portion of the total Hilbert space,
the particle undergoes a sort of Brownian motion, while the free Hamiltonian of
the spin array serves as a Wiener process.
No assumptions are made on the spectrum of the Hamiltonian of the spin system,
and no partial trace is computed over its states.
The mechanism of appearance of the stochastic process is discussed and
contrasted to other noteworthy examples in the literature. 
The links with van Hove's ``$\lambda^2 T$ limits are emphasized.

}

\end{titlepage}

\newpage


\setcounter{equation}{0}
\section{Introduction  }
\label{sec-introd}
\andy{intro}
  The derivation of a dissipative dynamics in quantum mechanics and
quantum field theory is a long-standing problem and is now being
widely investigated. A very important contribution to this issue
was given by van Hove \cite{vanHove}, who 
clarified the main features of a quantum dissipative dynamics and
was able to derive a master equation from the Schr\"odinger equation,
in an appropriate limit (his famous ``$\lambda^2T$" limit),
via the so-called ``diagonal singularity."
It is important to stress that van Hove derived a quantum dissipative dynamics
without making use of Pauli's random-phase approximation \cite{Pauli}.

Dissipation in quantum mechanics can emerge as a result of the interaction
between a particle and a macroscopic ``environment." 
Many other important contributions have been given to this issue, in the
attempt to set up a general physical framework. 
However, unlike in van Hove's case, a dissipative dynamics
is derived under some assumptions for the energy spectrum of the environment
system, by computing the partial trace over the states of the latter. This is
true for a whole class of interesting models, in which the environment
is schematized with a collection of harmonic oscillators \cite{FKM,CL}, and one
derives dissipative equations for the ``object" particle. 
It is worth observing that the dissipation constant in the Langevin-type
equation is not really derived from the underlying dynamics: It rather
appears as a consequence (although a reasonable one) of some assumtions on the 
spectrum of the Hamiltonian of the environment system.
Well-known peculiar properties of a ``quantum Langevin equation" are its 
nonmarkoffian random force \cite{FLOC} and colored noise \cite{MMN,Gardi2}.

In this paper we shall analyze the so-called ``AgBr" model \cite{CH,Bell}, that
has played an important role in the quantum measurement problem. This model is
relatively simple, yet extremely interesting from the physical point of view.
We shall base our discussion on a modified version \cite{NaPa3} 
of the above model, that is able 
to take into account energy-exchange processes. The modified AgBr model
provides an interesting nontrivial example of realization of van Hove's
diagonal singularity and displays the occurrence of an exponential regime at
all times in the weak-coupling, macroscopic limit \cite{NNP1}. In this way, a
door is open to investigate the occurrence of a dissipative dynamics and its
link with a quantum measurement process. 
No assumptions shall be made on the spectrum of the macroscopic system (that
plays the role of environment) in order to derive a dissipative dynamics. We
shall just solve the equations of motion, take a weak-coupling, macroscopic
limit, and obtain a Wiener process in a restricted portion of the Hilbert 
space of the total system. 
This technique is to be contrasted to the computation of a partial trace 
over the states of the macroscopic system, and will represent the main 
difference between the present work and other ones, based on partial tracing.

This work has interesting spinoffs for the quantum 
measurement problem \cite{von}.
It is commonly believed that a quantum measurement
occurs via a dephasing (decoherence) process
\cite{Zurek,MN,Zurenv}. Since 
`decoherence' is nothing but the disapperance of the off-diagonal
elements of the density matrix of the quantum system, and since
a system described by a diagonal density
matrix exhibits a purely stochastic behavior \cite{Wax},
one is led to expect
a connection between dissipation, irreversibility and a quantum
measurement process \cite{NNP1,Leggett,NNP2}.
In this paper, we shall concentrate on a particular aspect of the 
above-mentioned problems. Our main purpose is to derive a stochastic process 
from an underlying Hamiltonian dynamics and to clarify in which sense it is 
possible to identify a Wiener process for a dynamical variable of the 
microscopic system under investigation.

The plan of the paper is as follows.
We review the main properties of the modified AgBr Hamiltonian in Sec.\ 2,
focusing on those characteristics that hint at the presence of a stochastic 
process. In Sec.\ 3
we obtain the operators in the Heisenberg picture. 
{}Finally, we bring to light the stochastic process in Sec.\ 4, 
where we show that there is a Gaussian process of the Wiener type.
All these results are exact.
Section 5 is devoted to conclusions and comments.

\setcounter{equation}{0}
\section{Review of the AgBr Hamiltonian}
\label{sec-abha}
\andy{abha}
The modified AgBr Hamiltonian \cite{NaPa3} describes the interaction between an
ultrarelativistic particle $Q$ and a 1-dimensional array ($D$-system), made up
of $N$ spin-1/2 objects. The array can be viewed as a caricature of a linear
``photographic emulsion" of AgBr molecules, if one identifies the {\em down}
(ground) state of the spin with the undivided molecule and the {\em up} state
(whose excitation energy is $\hbar\omega$) with the dissociated molecule (Ag
and Br atoms). The particle and each molecule interact via a spin-flipping
local potential. The total Hamiltonian for the $Q+D$ system reads 
\andy{totham}
\beq
H = H_{0} + H',  \qquad
\qquad H_0 = H_{Q} + H_{D},
\label{eq:totham}
\eeq
where $H_{Q}$ and $H_{D}$,
the free Hamiltonians of the $Q$ particle and of the ``detector" $D$,
respectively, and the interaction Hamiltonian $H'$ are written as
\andy{H}
\barr
H_{Q} & = & c \hh{p},    \qquad
     H_{D}  =  \frac{1}{2}  \hbar  \omega
  \sum_{n=1}^{N}  \left( 1+\sigma_{3}^{(n)} \right) , \nonumber  \\
H' & = & \sum_{n=1}^{N} V(\hh{x}- x_n)
  \left[ \sigma_{+}^{(n)} \exp \left( -i \frac{\omega}{c}\hh{x}\right)+
\sigma_{-}^{(n)}\exp\left(+i\frac{\omega}{c}\hh{x}\right)\right].
\label{eq:H}
\earr
Here $\hh{p}$ is the momentum of the $Q$ particle, $\hh{x}$ its position,
$V$ a real potential, $x_n\; (n=1,...,N)$ the positions of the
scatterers in the array $(x_n>x_{n-1})$ and $\sigma_{i,\pm}^{(n)}$
the Pauli matrices acting on the $n$th site.
The above Hamiltonian has attracted the attention of several researchers
\cite{CHatt} due, in particular, to the presence
of the free Hamiltonian of the array $H_D$, which enables one to 
distinguish energetically the up and down states and makes
the energy-exchange processes between $Q$ and $D$ physically meaningful.
The original Hamiltonian \cite{CH} is reobtained in the $\omega=0$ limit.

Let us review the main results obtained from this model \cite{NaPa3,NNP1}.
The evolution operator in the interaction picture
\andy{evoop}
\beq
U_{I}(t,t') =e^{iH_{0}t/\hbar}e^{-iH(t-t')/\hbar}e^{-iH_{0}t'/\hbar}
=e^{-i\int_{t'}^{t}H'_{I}(t')dt'/\hbar},
\label{eq:evoop}
\eeq
where $H'_{I}(t)$ is the interaction Hamiltonian in the interaction picture, 
can be computed exactly as 
\andy{mwow}
\barr
U_I(t)&\equiv&U_I(t,0) =e^{iH_{0}t/\hbar}e^{-iHt/\hbar}\nonumber\\
&=&\prod_{n=1}^{N}\mbox{exp}\left[-\frac{i}{\hbar}\int_{0}^{t}dt'
V(\hh{x}+ct'-x_{n})\left(\sigma_{+}^{(n)}e^{-i\frac{\omega}{c}\hh{x}}+
\mbox{H.c.}\right)\right],
\label{eq:mwow}
\earr
and a straightforward calculation yields the $S$-matrix
\andy{scamm}
\beq
S^{[N]}=
\lim_{\stackrel{\scriptstyle t\rightarrow+\infty}
     {_{\scriptstyle t'\rightarrow-\infty}}}
U_I(t,t')
=\prod_{n=1}^{N}S_{(n)}:
\quad S_{(n)}=
\mbox{exp}\left[-i\frac{V_{0}\Omega}{\hbar c}
\mbox{\boldmath$\sigma$}^{(n)}\cdot\mbox{\boldmath$u$}
\right],
\label{eq:scamm}
\eeq
where $\mbox{\boldmath$u$}=[\cos(\omega x/c),\sin(\omega x/c),0]$ and 
$V_{0}\Omega\equiv\int_{-\infty}^{\infty}V(x)dx<\infty$.
The above expression enables us to define the ``spin-flip'' probability, i.e. 
the probability of dissociating one AgBr molecule:
\andy{prdis}
\beq
q\equiv\sin^{2}\left[\frac{V_{0}\Omega}{c\hbar}\right].
\label{eq:prdis}
\eeq
By defining
\andy{alde}
\beq
\alpha_{n}\equiv\alpha_{n}(\hh{x},t)
\equiv\int_{0}^{t}\frac{dt'}{\hbar} V(\hh{x}+ct'-x_{n})
\label{eq:alde}
\eeq
and
\andy{sixde}
\beq
\sigma_{\pm}^{(n)}(\hh{x})\equiv
\sigma_{\pm}^{(n)}e^{\mp i\frac{\omega}{c}\hh{x}}\,,
\label{eq:sixde}
\eeq
which satisfy, together with $\sigma_{3}^{(n)}$, the SU(2) algebra
\andy{su2}
\begin{eqnarray}
\left[\sigma_{3}^{(m)},\sigma_{\pm}^{(n)}(\hh{x})\right]
&=&\pm 2\delta_{mn}\sigma_{\pm}^{(n)}(\hh{x})\,,\nonumber\\
\left[\sigma_{-}^{(m)}(\hh{x}),\sigma_{+}^{(n)}(\hh{x})\right]
&=&-\delta_{mn}\sigma_{3}^{(n)}\,,
\label{eq:su2}
\end{eqnarray}
we can return to the Schr\"odinger picture by inverting 
Eq.\ (\ref{eq:mwow}). 
We disentangle the exponential in $U_{I}$ by making use 
of (\ref{eq:su2}) and obtain 
\andy{diseh}
\beq
e^{-iHt/\hbar}=e^{-iH_0t/\hbar}
\prod_{n=1}^{N}\left(
e^{-i\tan(\alpha_{n})\sigma_{+}^{(n)}(\hh{x})}
e^{-\ln\cos(\alpha_{n})\sigma_{3}^{(n)}}
e^{-i\tan(\alpha_{n})\sigma_{-}^{(n)}(\hh{x})}\right).
\label{eq:diseh}
\eeq
Let us concentrate our attention on the situation in which the $Q$ 
particle is initially located at $x'<x_{1}$, where $x_{1}$ is the position of 
the first scatterer in the linear array, and moves toward the array with 
speed $c$. 
The spin system is initially set in the ground state $|0\rangle_{N}$ 
of the free Hamiltonian $H_{D}$ (all spins down). 
The propagator is defined by
\andy{propaa}
\beq
G(x,x',t)\equiv
\langle x|\otimes\,_{N}\langle 0|
e^{-iHt/\hbar}
|0\rangle_{N}\otimes|x'\rangle\,.
\label{eq:propaa}
\eeq
If we place the spin array at the far right of the origin $(x_{1}>0)$ and 
consider the case in which the potential $V$ has a compact support and the $Q$ 
particle is initially located at the origin $x'=0$, we obtain
\andy{propgg2}
\beq
G(x,0,t)= \delta(x-ct)\prod_{n=1}^{N}\cos\tilde{\alpha}_{n}(t),
\qquad\tilde{\alpha}_{n}(t)\equiv
\int_{0}^{t}\frac{dt'}{\hbar} V(ct'-x_{n})\,.
\label{eq:propgg2}
\eeq
Note that, due to the choice of the free Hamiltonian $H_{Q}$, the $Q$ wave 
packet does not disperse and moves with constant speed $c$.
In this paper we shall exclusively consider the weak-coupling, macroscopic 
limit
\andy{rimba}
\beq
N\longrightarrow\infty\quad\mbox{and}\quad
q\simeq\left(\frac{V_{0}\Omega}{\hbar c}\right)^2=O(N^{-1}),
\label{eq:rimba}
\eeq
which is equivalent to the requirement that the total number of spin flips 
$\bar{n}=qN$ be finite in the macroscopic limit $N\to\infty$.
Notice that, if we set
\andy{limma0}
\beq
x_{n}=x_{1}+(n-1)d,\qquad L=x_{N}-x_{1}=(N-1)d,
\label{eq:limma0}
\eeq
and let $d/L\longrightarrow 0$ as $N\longrightarrow\infty$, a summation over
$n$ can be replaced by a definite integration according to 
\andy{regg-} 
\beq
q\sum_{n=1}^{N}f(x_{n})
\longrightarrow
\frac{\bar{n}}{L}\int_{x_{1}}^{x_{N}}dy f(y).
\label{eq:regg-}
\eeq
In this case, by making use of the Fermi-Yang approximation 
$V(y) = V_{0}\Omega\delta(y)$, (\ref{eq:propgg2}) becomes
\andy{prrfin}
\begin{eqnarray}
G&\propto&\mbox{exp}\left(\sum_{n=1}^{N}
\ln\left[\cos\int_{0}^{ct}dx\frac{V_{0}\Omega}{c\hbar}\delta(x-ct)
\right]\right)\nonumber\\
&\longrightarrow&\mbox{exp}\left(-\frac{\bar{n}}{2L}\left[
(ct-x_{1})\theta(x_{N}-ct)\theta(ct-x_{1})+L\theta(ct-x_{N})\right]\right)\,,
\label{eq:prrfin}
\end{eqnarray}
where the arrow will henceforth denote the weak-coupling, macroscopic limit
(\ref{eq:rimba}), (\ref{eq:regg-}).
The system attains an exponential regime as soon as the interaction starts:
Indeed, if $x_{1}<ct<x_{N}$,
\andy{reess}
\beq
G\propto\mbox{exp}\left(-\bar{n}\frac{(ct-x_{1})}{2L}\right).
\label{eq:reess}
\eeq
Notice the absence of the ``Gaussian" regime, characterized by a vanishing
derivative at $t=0$ \cite{stbeh,qze}, and of the power law at long times
\cite{Hell}.
This result is valid for the propagator (\ref{eq:propaa}), which involves 
position eigenstates of the $Q$ particle. If these are substituted by 
(normalizable) wave packets, small deviations from the exponential law appear 
at short times \cite{NNP2}, in agreement with general mathematical theorems 
\cite{stbeh,Hell}.

The result (\ref{eq:reess}) hints at the presence of a dissipative
process of some sort, at least in a restricted portion of the Hilbert space 
of the total ($Q+D$) system. Such a dissipative process was brought to light 
in Ref.\ \cite{BNNP}, where it was shown that a Wiener process appears in
the weak-coupling, macroscopic limit 
(\ref{eq:rimba}), (\ref{eq:regg-}).
In the following sections we shall derive all 
results (including ``border effects") in full generality, discuss their 
meaning and clarify in which sense and under which conditions it is possible to
identify the presence of a dissipative process. The important links with 
van Hove's $\lambda^2 T$ limit \cite{vanHove} will also be properly emphasized.

\setcounter{equation}{0}
\section{Heisenberg operators}
\label{sec-Heis}
\andy{Heis}

In order to bring to light the emergence of a dissipative 
process in the particle-detector interaction, 
it is essential to study the temporal evolution of all the 
operators involved in the interaction process.
It is therefore convenient to work in the Heisenberg picture.
{}First of all, notice that the total Hamiltonian of the system 
is constant in time 
\andy{hcost}
\begin{eqnarray}
& &H(t)\nonumber\\
&=&c\hh{p}(t)+
\frac{\hbar\omega}{2}\sum_{n}\left(1+\sigma_{3}^{(n)}(t)\right)
+ \sum_{n}V(\hh{x}(t)-x_{n})
\left[\sigma_{+}^{(n)}(t)e^{-i\frac{\omega}{c}\hh{x}(t)}+
\sigma_{-}^{(n)}(t)e^{i\frac{\omega}{c}\hh{x}(t)}\right]
\nonumber \\
&=&H(0)\,,
\label{eq:hcost}
\end{eqnarray}
where $H(0)\equiv H$ is the total Hamiltonian of the system in the 
Schr\"odinger picture.
Let us focus our attention on the free Hamiltonian of the $Q$
particle. From (\ref{eq:hcost}) we get
\andy{cpt}
\begin{eqnarray}
c\hh{p}(t)
&=& c\hh{p}(0)+\frac{\hbar\omega}{2}
\sum_{n}\left(\sigma_{3}^{(n)}-\sigma_{3}^{(n)}(t)\right)\nonumber\\
& &+ \sum_{n}\left\{V(\hh{x}-x_{n})
\left[\sigma_{+}^{(n)}e^{-i\frac{\omega}{c}\hh{x}}+ H.c.\right]\right.
\nonumber\\
& &\phantom{\sum_{n}V}\left. 
-V(\hh{x}(t)-x_{n})
\left[\sigma_{+}^{(n)}(t)e^{-i\frac{\omega}{c}\hh{x}(t)}+ H.c.\right]\right\} ,
\label{eq:cpt}
\end{eqnarray}
where $c\hh{p}(0)\equiv c\hh{p}$ and $\sigma_{i}^{(n)}\equiv 
\sigma_{i}^{(n)}(0)$ $(i=3,\pm)$ are operators in the Schr\"{o}dinger picture.
In order to solve Eq.\ (\ref{eq:cpt}), we need the explicit forms of the 
Heisenberg operators $\hh{x}(t)$, $\sigma_{3}^{(n)}(t)$ and 
$\sigma_{\pm}^{(n)}(t)$.
To this end, we shall make use of the disentanglement formula (\ref{eq:diseh}).

The calculation of the operator $\hh{x}(t)$ is straightforward and yields 
\andy{calxx}
\beq
\hh{x}(t)=e^{iHt/\hbar}\hh{x}e^{-iHt/\hbar}=\hh{x}+ct .
\label{eq:calxx}
\eeq
On the other hand, the evaluation of $\sigma_{i,\pm}^{(n)}(t)$ is more 
involved; let us first show in full how to calculate the operator 
$\sigma_{+}^{(n)}(t)$.  
With the help of (\ref{eq:diseh}) this operator can be rewritten as
\andy{cal+}
\begin{eqnarray}
\sigma_{+}^{(n)}(t)
&=&e^{iHt/\hbar}\sigma_{+}^{(n)}e^{-iH_{0}t/\hbar}\prod_{m}D_{m},
\label{eq:cal+}\\
D_{m}
&\equiv&e^{-i\tan(\alpha_{m})\sigma_{+}^{(m)}(\hh{x})}
        e^{-\ln\cos(\alpha_{m})\sigma_{3}^{(m)}}
        e^{-i\tan(\alpha_{m})\sigma_{-}^{(m)}(\hh{x})}.\nonumber
\end{eqnarray}
By observing that
\andy{cal+1}
\barr
& &e^{iHt/\hbar}\sigma_{+}^{(n)}e^{-iH_{0}t/\hbar}
  =e^{iHt/\hbar}e^{i\frac{\omega}{c}\hh{x}}\sigma_{+}^{(n)}(\hh{x})
   e^{-iH_{0}t/\hbar}\nonumber\\
&=&e^{i\frac{\omega}{c}(\hh{x}+ct)}e^{iHt/\hbar}\sigma_{+}^{(n)}(\hh{x})
   e^{-iH_{0}t/\hbar}
  =e^{i\frac{\omega}{c}(\hh{x}+ct)}e^{iHt/\hbar}
   e^{-iH_{0}t/\hbar}\sigma_{+}^{(n)}(\hh{x}),
\label{eq:cal+1}
\earr
Eq.\ (\ref{eq:cal+}) becomes 
\andy{cal+2}
\beq
e^{i\frac{\omega}{c}(\hh{x}+ct)}e^{iHt/\hbar}e^{-iH_{0}t/\hbar}
\left[\sigma_{+}^{(n)}(\hh{x})
e^{-i\tan(\alpha_{n})\sigma_{+}^{(n)}(\hh{x})}
e^{-\ln\cos(\alpha_{n})\sigma_{3}^{(n)}}
e^{-i\tan(\alpha_{n})\sigma_{-}^{(n)}(\hh{x})}\right]\prod_{m\neq n}D_{m}.
\label{eq:cal+2}
\eeq
We now evaluate the term in square brackets.
By making use of the formulas
\andy{3+,-+}
\beq
e^{a\sigma_{3}^{(n)}}\sigma_{+}^{(n)}e^{-a\sigma_{3}^{(n)}}
= e^{2a}\sigma_{+}^{(n)},\qquad
e^{b\sigma_{-}^{(n)}}\sigma_{+}^{(n)}e^{-b\sigma_{-}^{(n)}}
= \sigma_{+}^{(n)}-b\sigma_{3}^{(n)}+\frac{b^{2}}{2!}(-2\sigma_{-}^{(n)}),
\label{eq:3+,-+}
\eeq
we obtain 
\andy{cal+3}
\begin{eqnarray}
& &\sigma_{+}^{(n)}(\hh{x})
e^{-i\tan(\alpha_{n})\sigma_{+}^{(n)}(\hh{x})}
e^{-\ln\cos(\alpha_{n})\sigma_{3}^{(n)}}
e^{-i\tan(\alpha_{n})\sigma_{-}^{(n)}(\hh{x})}
\nonumber\\
&=&e^{-i\tan(\alpha_{n})\sigma_{+}^{(n)}(\hh{x})}
e^{-\ln\cos(\alpha_{n})\sigma_{3}^{(n)}}
e^{2\ln\cos(\alpha_{n})}
\sigma_{+}^{(n)}(\hh{x})
e^{-i\tan(\alpha_{n})\sigma_{-}^{(n)}(\hh{x})}
\nonumber\\
&=&e^{-i\tan(\alpha_{n})\sigma_{+}^{(n)}(\hh{x})}
e^{-\ln\cos(\alpha_{n})\sigma_{3}^{(n)}}
e^{-i\tan(\alpha_{n})\sigma_{-}^{(n)}(\hh{x})}\cos^{2}\alpha_{n}
\nonumber\\
& &\qquad\times
\left(\sigma_{+}^{(n)}(\hh{x})
-i\sigma_{3}^{(n)}\tan\alpha_{n}
-(i\tan\alpha_{n})^{2}\sigma_{-}(\hh{x})\right).
\nonumber\\
\label{eq:cal+3}
\end{eqnarray}
By substituting (\ref{eq:cal+3}) into Eq.\ (\ref{eq:cal+2}), we get
\andy{cal+4}
\begin{eqnarray}
\sigma_{+}^{(n)}(t)&=&
e^{i\frac{\omega}{c}(\hh{x}+ct)}e^{iHt/\hbar}
e^{-iH_{0}t/\hbar}\prod_{m}
e^{-i\tan(\alpha_{m})\sigma_{+}^{(m)}(\hh{x})}
e^{-\ln\cos(\alpha_{m})\sigma_{3}^{(m)}}
e^{-i\tan(\alpha_{m})\sigma_{-}^{(m)}(\hh{x})}
\nonumber\\
& &\times\left(\sigma_{+}^{(n)}(\hh{x})\cos^{2}\alpha_{n}
-\frac{i}{2}\sigma_{3}^{(n)}\sin 2\alpha_{n}
+\sigma_{-}^{(n)}(\hh{x})\sin^{2}\alpha_{n}\right)
\nonumber\\
&=&e^{i\frac{\omega}{c}(\hh{x}+ct)}
\left(\sigma_{+}^{(n)}(\hh{x})\cos^{2}\alpha_{n}
-\frac{i}{2}\sigma_{3}^{(n)}\sin 2\alpha_{n}
+\sigma_{-}^{(n)}(\hh{x})\sin^{2}\alpha_{n}\right) .
\label{eq:cal+4}
\end{eqnarray}
The Hermitian conjugate of (\ref{eq:cal+4}) yields
\andy{cal-}
\beq
\sigma_{-}^{(n)}(t)
=e^{-i\frac{\omega}{c}(\hh{x}+ct)}
\left(\sigma_{-}^{(n)}(\hh{x})\cos^{2}\alpha_{n}
+\frac{i}{2}\sigma_{3}^{(n)}\sin 2\alpha_{n}
+\sigma_{+}^{(n)}(\hh{x})\sin^{2}\alpha_{n}\right) 
\label{eq:cal-}
\eeq
and by using an analogous procedure, together with the formula
\andy{+3}
\beq
e^{b\sigma_{+}^{(n)}}\sigma_{3}^{(n)}e^{-b\sigma_{+}^{(n)}}
= \left(\sigma_{3}^{(n)}-2b\sigma_{+}^{(n)}\right) 
\label{eq:+3}
\eeq
and its Hermitian conjugate, we get
\andy{cal3}
\beq
\sigma_{3}^{(n)}(t)=\sigma_{3}^{(n)}\cos 2\alpha_{n}
-i\left(\sigma_{+}^{(n)}(\hh{x})-\sigma_{-}^{(n)}(\hh{x})\right)
\sin 2\alpha_{n} .
\label{eq:cal3}
\eeq
The results (\ref{eq:cal+4}), (\ref{eq:cal-}) and (\ref{eq:cal3}) 
and the relation 
\andy{osst0}
\beq
\sigma_{\pm}^{(n)}(t)e^{\mp i\frac{\omega}{c}(\hh{x}+ct)}=
\sigma_{\pm}^{(n)}e^{\mp i\frac{\omega}{c}\hh{x}}=\sigma_{\pm}^{(n)}(\hh{x}),
\label{eq:osst0}
\eeq
lead us to the final expression 
\andy{opcpt}
\begin{eqnarray}
c\hh{p}(t)&=&c\hh{p}(0)+
\hbar\omega\sum_{n}\sigma_{3}^{(n)}\sin^{2}\alpha_{n}(\hh{x},t)+ 
i\frac{\hbar\omega}{2}\sum_{n}\left[\left(\sigma_{+}^{(n)}(\hh{x})
                                          -\sigma_{-}^{(n)}(\hh{x})\right)
\sin 2\alpha_{n}(\hh{x},t)\right]   \nonumber \\
  & & + \sum_{n}\left[V(\hh{x}-x_{n})-V(\hh{x}+ct-x_{n})\right]
\left(\sigma_{+}^{(n)}(\hh{x})+\sigma_{-}^{(n)}(\hh{x})\right). 
\label{eq:opcpt}
\end{eqnarray}
All the above results are {\em exact}.

\setcounter{equation}{0}
\section{The stochastic process}
\label{sec-stopro}
\andy{stopro}
Having found the explicit (and exact) expressions for the quantum operators 
of our system in the Heisenberg picture, we search for 
the {\em stochastic process} that is at the origin of the exponential 
decay (\ref{eq:reess}) of the propagator (\ref{eq:propaa}).
\subsection{Expectation value of the $Q$ particle Hamiltonian}
\label{subsec-vall}
\andy{vall}
By making use of the explicit expression (\ref{eq:opcpt}) for the momentum 
operator of the $Q$ particle in the Heisenberg picture, we can easily compute 
its expectation value in the state 
\andy{fonst}
\beq
|\psi,0\rangle_{N}\equiv|\psi\rangle\otimes|0\rangle_{N}=
\int dx\psi(x)|x\rangle\otimes |0\rangle_{N},\qquad
\int_{-\infty}^{+\infty}dx|\psi(x)|^{2}=1.
\label{eq:fonst}
\eeq
The choice of such an uncorrelated initial state is physically consistent:
Indeed we suppose that at initial time $t=0$ the particle $Q$ is well 
outside the detector $D$ and moves toward it with constant speed $c$. 
It will be clear in the following that the choice of the ground state 
$|0\rangle_{N}$ ($N$ spins down) as the initial $D$ state is essential in 
deriving a stochastic process.
{}For convenience we choose $\psi$ to be symmetrically distributed around the 
origin and with a compact support.
Possible choices for $\psi$ are 
\andy{tet}
\beq
\psi(x)=(2a)^{-1/2}\theta (a-|x|)e^{ip_{0}x/\hbar}\,,
\label{eq:tet}
\eeq
or a Gaussian wave packet 
\andy{gauss}
\beq
\psi(x)=\left(\frac{1}{2\pi a^{2}}\right)^{\frac{1}{4}}
        e^{-\frac{x^{2}}{4a^{2}} + i\frac{p_{0}x}{\hbar}},
\label{eq:gauss}
\eeq
truncated for, say, $|x|>a$. Notice again that, owing to (\ref{eq:calxx}), the
wave packet does not disperse. We define: 
\andy{medcpt1}
\beq
\langle c\hh{p}(t)\rangle
\equiv{}_{N}\langle0,\psi|c\hh{p}(t)|\psi,0\rangle_{N}
\label{eq:medcpt1}
\eeq
and, from Eq.\ (\ref{eq:opcpt}) we obtain
\andy{medcpt2}
\begin{eqnarray}
\langle c\hh{p}(t)\rangle
&=& \,_{N}\langle 0,\psi| c\hh{p}(0) |\psi ,0\rangle_{N} +\hbar\omega\,_{N}
\langle 0,\psi |\sum_{n}\sigma_{3}^{(n)}\sin^{2}\alpha_{n}(\hh{x},t)
|\psi ,0\rangle_{N} \nonumber \\
& &+i\frac{\hbar\omega}{2}\,_{N}\langle 0,\psi |\sum_{n}
\left(\sigma_{+}^{(n)}(\hh{x})-\sigma_{-}^{(n)}(\hh{x})\right)
\sin 2\alpha_{n}(\hh{x},t)|\psi ,0\rangle_{N} \nonumber \\
& &+\,_{N}\langle 0,\psi |\sum_{n}\left[V(\hh{x}-x_{n})- V(\hh{x}+ ct-x_{n})
\right]\left(\sigma_{+}^{(n)}(\hh{x})+\sigma_{-}^{(n)}(\hh{x})\right)
|\psi ,0\rangle_{N}. \nonumber\\
\label{eq:medcpt2} 
\end{eqnarray}
It is easy to see that the last two terms in (\ref{eq:medcpt2}) give 
vanishing contributions. 
Let us now compute the contributions of the first and the 
second terms.
The evaluation of the first term is straightforward: It 
is nothing but the initial energy of the $Q$ particle
\andy{primo}
\begin{eqnarray}
\,_{N}\langle 0,\psi| c\hh{p}(0) |\psi ,0\rangle_{N}
&=& \langle\psi |c\hh{p}(0) |\psi\rangle = \int dp'\langle\psi 
|p'\rangle\langle p'| c\hh{p} |\psi\rangle \nonumber\\
&=& \int dp' |\tilde{\psi}(p')|^{2}cp' = cp_{0}. 
\label{eq:primo}
\end{eqnarray}
On the other hand, the calculation of the second term 
\andy{secondo}
\begin{eqnarray}
& &\hbar\omega\,_{N}\langle 0,\psi |\sum_{n}\sigma_{3}^{(n)}
\sin^{2}\alpha_{n}(\hh{x},t)|\psi ,0\rangle_{N} = -\hbar\omega\langle\psi |   
 \sum_{n}\sin^{2}\alpha_{n}(\hh{x},t) |\psi\rangle  \nonumber\\
& &\qquad = -\hbar\omega\int dx |\psi (x)|^{2}\sum_{n}\sin^{2}\left[
\frac{1}{\hbar}\int_{0}^{t} dt'V( x + ct'- x_{n})\right]
\label{eq:secondo}
\end{eqnarray}
is more involved:
We can use, without loss of generality, the Fermi-Yang approximation 
$V(y) = V_{0}\Omega\delta(y)$, and obtain
\andy{eq:mod2}
\begin{eqnarray}
&&-\hbar\omega\int dx |\psi (x)|^{2}
\sum _{n}\sin^{2}\left[
\frac{V_{0}\Omega}{c\hbar}\int_{x-x_{n}}^{x+ct-x_{n}} dy \delta (y)\right]  
\nonumber\\
&=&-\hbar\omega\int dx |\psi (x)|^{2}
\sum _{n}\sin^{2}\left[
\frac{V_{0}\Omega}{c\hbar}\theta ( x + ct - x_{n})\theta ( x_{n}- x)\right].
\label{eq:mod2}
\end{eqnarray}
If we require
\andy{stree}
\beq
a\ll x_{1}\quad\mbox{and}\quad a\ll L=x_{N}-x_{1}, 
\label{eq:stree}
\eeq
we easily get 
\andy{mod3}
\beq
 -\hbar\omega\int dx |\psi (x)|^{2}\sum _{n}\sin^{2}
\left[\frac{V_{0}\Omega}{c\hbar} \theta ( x + ct - x_{n})\right],
\label{eq:mod3}
\eeq
since any spin is located at the far right of the initial wave packet 
$\psi(x)$, whose finite support is $[-a,a]$, and therefore the inequality
$x(<a)\ll x_n$ holds for the integration variable $x$. 
In the weak-coupling, macroscopic limit (\ref{eq:rimba}), the summation over 
$n$ can be replaced with an integration as in (\ref{eq:regg-}) and the 
above quantity is further reduced to 
\andy{mod6}
\begin{eqnarray}   
&&-\hbar\omega\frac{\bar{n}}{L}\int dx |\psi (x)|^{2}
\int_{x+ct-x_{N}}^{x+ct-x_{1}}dz \theta (z)\nonumber\\ 
&=&-\hbar\omega\frac{\bar{n}}{L}\int dx |\psi (x)|^{2}
   \left[(x+ct-x_{1})\theta (x+ct-x_{1})\theta (x_{N}-ct-x)\right.\nonumber\\
\noalign{\vspace{-2pt}}
& &\phantom{-\hbar\omega\frac{\bar{n}}{L}\int dx |\psi (x)|^{2}}
   +\left.(x_{N}-x_{1})\theta (x+ct-x_{N})\right]. 
\label{eq:mod6}
\end{eqnarray}
Thus, if we restrict our attention to the situation in which $Q$ has not gone 
through $D$, i.e.\ $ct<x_N$, we finally obtain
\andy{mod8}
\beq
\hbar\omega\,_{N}\langle 0,\psi |
\sum_{n}\sigma_{3}^{(n)}\sin^{2}\alpha_{n}(\hh{x},t)
|\psi ,0\rangle_{N} \longrightarrow
 -\hbar\omega\frac{\bar{n}}{L}(ct - x_{1}) + \mbox{b.e.}.
\label{eq:mod8}
\eeq
Notice that ``b.e." is a shorthand notation for ``border effects'', namely 
terms appearing only when $|ct-x_{1}|,|ct-x_{N}|\leq a$, whose explicit 
expression is easily computed to be 
\andy{boef}
\beq
-\hbar\omega\frac{\bar{n}}{L}\times\left\{
\begin{array}{l}
(ct-x_1+a)^2/4a\\
\noalign{\vspace{6pt}}
L-(x_N+a-ct)^2/4a
\end{array}
\right.
\begin{array}{l}
\quad\mbox{if}\quad x_1-a\leq ct\leq x_1+a,\\
\noalign{\vspace{6pt}}
\quad\mbox{if}\quad x_N-a\leq ct\leq x_N+a,
\end{array}
\label{eq:boef}
\eeq
corresponding respectively to the situations in which $Q$ is entering $D$
and $Q$ is going out of $D$.

Having obtained the explicit expressions (\ref{eq:primo}) and (\ref{eq:mod8}) 
for the first and the second terms in $\langle c\hh{p}(t)\rangle$ [see 
(\ref{eq:medcpt2})], we reach the following {\em exact} expression
\andy{medcpt3}
\beq
\langle c\hh{p}(t)\rangle \equiv\,_{N}\langle 0,\psi | c\hh{p}(t) |\psi ,0
\rangle_{N} \longrightarrow cp_{0}- 
\hbar\omega\frac{\bar{n}}{L}(ct - x_{1}) + \mbox{b.e.}.
\label{eq:medcpt3}
\eeq
In conclusion, the energy of the $Q$ particle (when $Q$ is {\em inside} $D$) 
decreases linearly with respect to $t$.
It is worth mentioning that what is seen here is an 
energy-dissipative process:
If $\omega$ were set equal to zero (as in the original AgBr 
model \cite{CH,Bell}), we could not have found such a process.

\subsection{Correlation functions of the spin-array Hamiltonian}
\label{subsec-vasp}
\andy{vasp}
In order to clarify the stochastic nature of the system, let us calculate the 
correlation functions of the spin-array Hamiltonian $H_D$. 

Consider the operator 
$\triangle H_D(t)=H_D(t)-H_D$, which represents the 
energy stored in the detector between time $0$ and $t$.
By using Eqs.\ (\ref{eq:cal3}), one obtains
\andy{H't}
\begin{eqnarray}
\triangle H_D(t)
&=&\sum_{n}\frac{\hbar\omega}{2}( \sigma_{3}^{(n)}(t)- \sigma_{3}^{(n)})
\nonumber\\ 
&=&-\left[\sum_{n}\hbar\omega\sigma_{3}^{(n)}\sin^{2}\alpha_{n}(\hh{x},t) 
+\sum_{n}i\frac{\hbar\omega}{2}
\left(\sigma_{+}^{(n)}(\hh{x})-\sigma_{-}^{(n)}(\hh{x})\right)
 \sin2\alpha_{n}(\hh{x},t)\right].\nonumber\\
&&
\label{eq:H't}
\end{eqnarray}
It is easy to see that when the $Q$ particle is inside $D$, by making use of
(\ref{eq:mod8}), the expectation value of this operator reads 
\andy{medH'}
\beq
\langle\triangle H_D(t)\rangle = 
\,_{N}\langle0,\psi | \triangle H_D(t) |\psi ,0\rangle_{N} \longrightarrow
\hbar\omega\frac{\bar{n}}{L}( ct - x_{1})+\mbox{b.e.},
\label{eq:medH'}
\eeq
in agreement with Eq.\ (\ref{eq:medcpt3}): The energy lost by $Q$ is 
stored in $D$.

Next we turn our attention to the two-time correlation function, defined by
\andy{corr}
\beq
\langle\triangle H_D(t_{1})\triangle H_D(t_{2})\rangle\equiv
\,_{N}\langle 0,\psi |
\triangle H_D(t_{1})\triangle H_D(t_{2})|\psi ,0\rangle_{N}.
\label{Eq:corr}
\eeq
Its explicit form, by Eq.\ (\ref{eq:H't}), is 
\andy{corr1}
\begin{eqnarray}
& &\langle\triangle H_D(t_{1})\triangle H_D(t_{2})\rangle 
\nonumber \\
&=&\,_{N}\langle 0,\psi |\left(\sum_{n}\hbar\omega\sigma_{3}^{(n)}
   \sin^{2}\alpha_{n}(\hh{x},t_{1})\right)
   \left(\sum_{m}\hbar\omega\sigma_{3}^{(m)}
   \sin^{2}\alpha_{m}(\hh{x},t_{2})\right)|\psi ,0\rangle_{N} 
\nonumber\\
& &+\left(i\frac{\hbar\omega}{2}\right)^{2}\,_{N}\langle 0,\psi|
   \left[\sum_{n}\left(\sigma_{+}^{(n)}(\hh{x})-\sigma_{-}^{(n)}(\hh{x})\right)
    \sin 2\alpha_{n}(\hh{x},t_{1})\right]
\nonumber\\
& &\phantom{+\left(i\frac{\hbar\omega}{2}\right)^{2}\,_{N}\langle 0,\psi|}
   \times
   \left[\sum_{m}\left(\sigma_{+}^{(m)}(\hh{x})-\sigma_{-}^{(m)}(\hh{x})\right)
    \sin2\alpha_{m}(\hh{x},t_{2})\right]|\psi,0\rangle_{N} \nonumber\\
& &+\,\mbox{vanishing terms.} 
\label{eq:corr1}
\end{eqnarray}
The first term yields
\andy{primocorr}
\begin{eqnarray}
& &\!\!\!\!_{N}\langle0,\psi|
\left(\sum_{n}\hbar\omega\sigma_{3}^{(n)}\sin^{2}\alpha_{n}(\hh{x},t_{1})
\right)\left(
\sum_{m}\hbar\omega\sigma_{3}^{(m)}\sin^{2}\alpha_{m}(\hh{x},t_{2})\right)
|\psi,0\rangle_{N}
\nonumber\\
&=&(\hbar\omega)^{2}\int dx |\psi (x)|^{2}\left[\sum_{n}
\sin^{2}\left(\frac{1}{\hbar}\int_{0}^{t_{1}}dt V(x+ct-x_{n})\right)\right]
\nonumber\\
& &\phantom{(\hbar\omega)^{2}\int dx |\psi (x)|^{2}}
   \times\left[\sum_{m}\sin^{2}\left(\frac{1}{\hbar}\int_{0}^{t_{2}} dt' 
   V(x + ct' - x_{m})\right)\right],
\label{eq:primocorr}
\end{eqnarray} 
which is written, in the Fermi-Yang approximation $V(x) =V_{0}\Omega
\delta(x)$, as
\andy{primocorr3}
\begin{eqnarray}
&&(\hbar\omega)^2\int dx|\psi(x)|^2\left[\sum_n\sin^{2}\left(\frac{V_{0}
  \Omega}{c\hbar}\theta(x+ct_1-x_n)\theta(x_n-x)\right)\right]
\nonumber\\
&&\phantom{(\hbar\omega)^2\int dx|\psi(x)|^2}
  \times\left[\sum_m\sin^2\left(\frac{V_0\Omega}{c\hbar}\theta(x+ct_2-x_m)
  \theta (x_m-x)\right)\right].
\label{eq:primocorr3}
\end{eqnarray}
The weak-coupling, macroscopic limit (\ref{eq:rimba}), 
together with the continuum ansatz (\ref{eq:regg-}), reduces this quantity to
\andy{primocorr4}
\begin{eqnarray}
& &\left(\hbar\omega\frac{\bar{n}}{L}\right)^{2}\int dx|\psi(x)|^{2}
\left[(x+ct_{1}-x_{1})\theta (x+ct_{1}-x_{1})
       \theta(x_{N}-ct_{1}-x)\right.\nonumber\\
\noalign{\vspace{-10pt}}
& &\phantom{
    \left(\hbar\omega\frac{\bar{n}}{L}\right)^{2}\int dx|\psi(x)|^{2}
    \left[(x+ct_{1}-x_{1})\theta(x+ct_{1}-x_{1})\right.}
   \left.+L\theta(x+ct_{1}-x_{N})\right]\nonumber\\
\noalign{\vspace{-10pt}}
& &\phantom{
    \left(\hbar\omega\frac{\bar{n}}{L}\right)^{2}\int dx|\psi(x)|^{2}}
   \times\left[(x+ct_{2}-x_{1})\theta (x+ct_{2}-x_{1})\theta(x_{N}-ct_{2}-x)
   \right.\nonumber\\
\noalign{\vspace{-10pt}}
& &\phantom{
    \left(\hbar\omega\frac{\bar{n}}{L}\right)^{2}\int dx|\psi(x)|^{2}
    \left[(x+ct_{1}-x_{1})\theta (x+ct_{1}-x_{1})\right.}
   +\left.L\theta(x+ct_{2}-x_{N})\right].\nonumber\\
& &
\label{eq:primocorr4}
\end{eqnarray}
{}Finally, by focusing our attention on the situation in which the $Q$ 
particle is {\em inside} $D$, we obtain
\andy{primocorr6}
\begin{eqnarray}
& &\left(\hbar\omega\frac{\bar{n}}{L}\right)^{2}\int dx |\psi (x)|^{2}
(x+ct_{1}-x_{1})(x+ct_{2}-x_{1})\nonumber\\
& &=\left(\hbar\omega\frac{\bar{n}}{L}\right)^{2}
    (ct_{1}-x_{1})(ct_{2}-x_{1})+\mbox{b.e.}
    \quad\mbox{(for $x_1<ct_{1,2}<x_N$).}
\label{eq:primocorr6}
\end{eqnarray}
Let us now calculate the second term in Eq.\ (\ref{eq:corr1}), the 
nonvanishing term of which reads 
\andy{secondocorr}
\beq
\left(\frac{\hbar\omega}{2}\right)^2\int dx|\psi(x)|^2
\sum_{n}\sin2\alpha_{n}(x,t_1)\sin2\alpha_n(x,t_2).
\label{eq:secondocorr}
\eeq
In the usual weak-coupling, macroscopic limit (\ref{eq:rimba}), 
we obtain, after a little manipulation,
\andy{secondocorr3}
\begin{eqnarray}
&&
 (\hbar\omega)^2\int dx|\psi(x)|^2\frac{\bar n}{L}\int_{x_1}^{x_N}dy
 \theta(x+ct_1-y)\theta(x+ct_2-y)\nonumber\\
&=&
(\hbar\omega)^2\frac{\bar n}{L}[c\min(t_1,t_2)-x_1]+\mbox{b.e.},
 \quad\mbox{(for $x_1<ct_{1,2}<x_N$).}
\label{eq:secondocorr3}
\end{eqnarray} 

In conclusion, we are led to the following expression for the two-time
correlation function 
\andy{finn}
\beq
\langle\triangle H_D(t_{1})\triangle H_D(t_{2})\rangle
\longrightarrow
\left( \hbar\omega\frac{\bar{n}}{L}\right)^{2}
(ct_{1}-x_{1})(ct_{2}-x_{1})+
(\hbar\omega)^{2}\frac{\bar{n}}{L}[c\min (t_{1},t_{2})-x_{1}]+\mbox{b.e.},
\nonumber\\
\label{eq:finn}
\eeq
which is valid when the $Q$ particle is inside $D$.
The border effects will be discussed in the Appendix.

\subsection{The Wiener process}
\label{sec-hammod}
\andy{hammod}
The results we have obtained so far, (\ref{eq:medH'}) and (\ref{eq:finn}), 
look quite interesting: Introduce the operator 
\andy{sigg}
\beq
\hh{\Sigma}(t) \equiv \triangle H_D(t)-\langle\triangle H_D(t)\rangle,
\label{eq:sigg}
\eeq
where the expectation value is to be evaluated on the state 
spanned by $|\psi,0\rangle_N\equiv|\psi\rangle\otimes|0\rangle_N$.
Then, by Eqs.\ (\ref{eq:medH'}) and (\ref{eq:finn}),
we easily show the following properties 
\andy{siggmed,siggcorr}
\barr
\langle\hh{\Sigma}(t)\rangle & = & 0 
\label{eq:siggmed} \\
\noalign{\vspace{6pt}}
\langle\hh{\Sigma}(t_{1})\hh{\Sigma}(t_{2})\rangle&=& 
\langle\triangle H_D(t_{1})\triangle H_D(t_{2})\rangle-
\langle\triangle H_D(t_{1})\rangle\langle\triangle H_D(t_{2})
\rangle\nonumber\\ 
& \longrightarrow &(\hbar\omega)^{2}\frac{\bar{n}}{L}[c\min(t_{1},t_{2})-x_{1}]
+\mbox{b.e.},
\label{eq:siggcorr}
\end{eqnarray}
valid in the restricted state-space
spanned by $|\psi,0\rangle_N$.
These properties remind us of the characteristics of a {\em Wiener 
stochastic process} \cite{Wax}, i.e., a Gaussian process with a variance 
proportional to $\min(t,t')$, since the second relation (\ref{eq:siggcorr})
can be rewritten as
\andy{siggcorr'}
\beq
\langle\hh{\Sigma}(t_{1})\hh{\Sigma}(t_{2})\rangle
\longrightarrow(\hbar\omega)^{2}\frac{c\bar{n}}{L}\min(\tau_{1},\tau_{2})
+\mbox{b.e.},
\label{eq:siggcorr'}
\eeq
in terms of an ``interaction time" $\tau_{1,2}\equiv t_{1,2}-x_1/c$.
As a matter of fact, we can prove that the operator $\hh{\Sigma}(t)$ really 
serves as a Wiener process {\em in the restricted Hilbert space spanned by} 
$|\psi, 0\rangle_{N}$, in the weak-coupling, macroscopic limit.
To this end, we must show that the process is Gaussian, namely
we must prove that the correlation functions of any order can be written as 
a sum of products of two-time correlation functions over all possible 
combinations.
This will be done in Sec.\ \ref{subsec-calcorn}.

\subsection{Characteristic functional}
\label{subsec-calcorn}
\andy{calcorn}
In order to demonstrate in full generality the Gaussian 
property of the process, let us consider the characteristic functional 
\andy{funcar}
\beq
\phi[\beta]\equiv
\langle e^{\int dt\beta(t)\hh{\Sigma}(t)}\rangle,
\label{eq:funcar}
\eeq
which is subject to the normalization condition 
\andy{concar}
\beq
\phi[0]=\langle1\rangle
={}_N\langle0,\psi|\psi,0\rangle_N=1.
\label{eq:concar}
\eeq
We know that the characteristic functional is the generating functional of 
correlation functions
\andy{collegg}
\beq
\langle\hh{\Sigma}(t_1)\hh{\Sigma}(t_2)\cdots\hh{\Sigma}(t_n)\rangle
=\left.\frac{\delta^n\phi[\beta]}{\delta\beta(t_1)\cdots\delta\beta(t_n)}
 \right|_{\beta=0},
\label{eq:collegg}
\eeq
and that Gaussian processes are characterized by Gaussian 
characteristic functionals.

By making use of (\ref{eq:sigg}) together with (\ref{eq:H't}) and 
(\ref{eq:medH'}), we find
\andy{cal1}
\begin{eqnarray}
\phi[\beta]
&=&\int dx|\psi(x)|^2\,_{N}\langle0|\prod_n
   e^{-\hbar\omega\int dt\beta(t)\left[\sigma_3^{(n)}\sin^2\alpha_n(x,t)
   +\frac{i}{2}\left(\sigma_+^{(n)}(x)-\sigma_-^{(n)}(x)\right)
   \sin2\alpha_{n}(x,t)\right]}|0\rangle_{N}\nonumber\\
& &\times e^{-\int dt\beta(t) \langle\triangle H_D(t)\rangle},
\label{eq:cal1}
\end{eqnarray}
Let us now focus our attention on the factor
\andy{calf}
\begin{eqnarray}
e^{-\hbar\omega\int dt\beta(t)\left[\sigma_3^{(n)}\sin^{2}\alpha_{n}(x,t)
+\frac{i}{2}\left(\sigma_+^{(n)}(x)-\sigma_-^{(n)}(x)\right)
\sin2\alpha_{n}(x,t)\right]}
&=&e^{a_n\left[b_n\sigma_3^{(n)}+i\left(\sigma_+^{(n)}(x)-\sigma_-^{(n)}(x)
   \right)\right]}\nonumber\\
&\equiv&f(a_n,b_n),
\label{eq:calf}
\end{eqnarray}
where we have introduced the quantities $a_n,b_n\,\in\Re$:
\andy{ann,bnn}
\barr
a_{n}&\equiv&-\frac{\hbar\omega}{2}\int dt\beta(t)\sin2\alpha_{n}(x,t),
\label{eq:ann}\\
a_{n}b_{n}&\equiv&-\hbar\omega\int dt\beta(t)\sin^{2}\alpha_{n}(x,t).
\label{eq:bnn}
\earr
We try a disentanglement of $f$ in the following form 
\andy{calf4}
\beq
f(a,b)=
e^{iX(a,b)\sigma_{+}(x)}e^{Y(a,b)\sigma_{3}}
e^{-iX(a,b)\sigma_{-}(x)},
\label{eq:calf4}
\eeq
where, for the moment, the index $n$ has been suppressed for the sake of
simplicity and the functions $X(a,b),Y(a,b)\in\Re$ are to be determined later. 
The determination of $X$ and $Y$ is straightforward but somewhat involved.
Differentiation of $f$, in (\ref{eq:calf}), w.r.t.\ $a$ yields
\andy{delfdela}
\beq
\frac{\partial f(a,b)}{\partial a}
=\left[b\sigma_3+i\left(\sigma_+(x)-\sigma_-(x)\right)\right]f(a,b),
\label{eq:delfdela}
\eeq
while the disentangled form of $f$ in
(\ref{eq:calf4}) implies that the same quantity is to be equated with
\andy{parr}
\beq
\left[
i\frac{\partial X}{\partial a}\sigma_{+}(x)+
\frac{\partial Y}{\partial a}e^{iX\sigma_{+}(x)}\sigma_{3}
e^{-iX\sigma_{+}(x)}
-i\frac{\partial X}{\partial a}e^{iX\sigma_{+}(x)}
e^{Y\sigma_{3}}\sigma_{-}(x)e^{-Y\sigma_{3}}
e^{-iX\sigma_{+}(x)}\right]f(a,b).
\label{eq:parr}
\eeq
The calculation of the terms in square brackets is simple:
The second term is calculated to be
\andy{oss1}
\beq
e^{iX\sigma_{+}(x)}\sigma_{3}e^{-iX\sigma_{+}(x)}
=\sigma_{3}-iX\left[\sigma_{3},\sigma_{+}(x)\right]
=\sigma_{3}-2iX\sigma_{+}(x),
\label{eq:oss1}
\eeq
while the third term becomes
\andy{oss2}
\begin{eqnarray}
e^{iX\sigma_{+}(x)}e^{Y\sigma_{3}}\sigma_{-}(x)
e^{-Y\sigma_{3}}e^{-iX\sigma_{+}(x)}
&=&e^{-2Y}e^{iX\sigma_{+}(x)}\sigma_{-}(x)e^{-iX\sigma_{+}(x)}
\nonumber\\
&=&e^{-2Y}\left(\sigma_{-}(x)+ iX\sigma_{3}+X^{2}\sigma_{+}(x)\right).
\label{eq:oss2}
\end{eqnarray}
Therefore we have the equality
\andy{parr2}
\begin{eqnarray}
&&b\sigma_{3}+ i\left(\sigma_{+}(x)-\sigma_{-}(x)\right)
\nonumber\\
&=&
i\frac{\partial X}{\partial a}\sigma_{+}(x)+
\frac{\partial Y}{\partial a}\left(\sigma_{3}-iX2\sigma_{+}(x)\right)
-i\frac{\partial X}{\partial a}e^{-2Y}\left(\sigma_{-}(x)+ 
iX\sigma_{3}+ X^{2}\sigma_{+}(x)\right),\nonumber\\
&&
\label{eq:parr2}
\end{eqnarray}
from which we obtain the following set of differential equations:
\andy{sistparr1,2,3}
\begin{eqnarray}
b&=&\frac{\partial Y}{\partial a}+ X\frac{\partial X}{\partial a}e^{-2Y},\\
\label{eq:sistparr1}
1&=&\frac{\partial X}{\partial a}- 2X\frac{\partial Y}{\partial a}
-X^{2}\frac{\partial X}{\partial a}e^{-2Y},\\
\label{eq:sistparr2}
1&=&\frac{\partial X}{\partial a}e^{-2Y}.
\label{eq:sistparr3}
\end{eqnarray}
We can easily solve these equations since they are equaivalent to the 
following two equations
\andy{parrX,parrY}
\barr
\frac{\partial X}{\partial a} &=& 1 + 2bX - X^{2} 
\label{eq:parrX} \\
\frac{\partial Y}{\partial a} &=& b - X.
\label{eq:parrY}
\earr
Under the initial condition $X(0,b)=0$, the solution of (\ref{eq:parrX}) is 
readily obtained
\andy{xab}
\beq
X(a,b)=\frac{e^{2a\sqrt{b^{2}+1}}-1}
{(\sqrt{b^{2}+1}+b)+(\sqrt{b^{2}+1}-b)e^{2a\sqrt{b^{2}+1}}}.
\label{eq:xab}
\eeq
Then the function $Y$ is calculated, either from (\ref{eq:parrY}) or 
(\ref{eq:sistparr3}), to be
\andy{yab}
\beq
Y(a,b)= a\sqrt{b^{2}+1}+\ln\left[\frac
{2\sqrt{b^{2}+1}}{(\sqrt{b^{2}+1}+b)+(\sqrt{b^{2}+1}-b)e^{2a\sqrt{b^{2}+1}}}
\right].
\label{eq:yab}
\eeq
By plugging the solutions (\ref{eq:xab})-(\ref{eq:yab}) into the 
disentanglement formula (\ref{eq:calf}) we can evaluate the 
characteristic functional
\andy{cal2}
\begin{eqnarray}
\phi[\beta]&=&
\int dx |\psi (x)|^{2}\,_{N}\langle 0|\prod_{n}
e^{iX(a_{n},b_{n})\sigma_{+}^{(n)}(x)}
e^{Y(a_{n},b_{n})\sigma_{3}^{(n)}}
e^{-iX(a_{n},b_{n})\sigma_{-}^{(n)}(x)}|0\rangle_{N}
\nonumber\\
& &\times e^{-\int dt\beta(t) \langle\triangle H_D(t)\rangle} \nonumber\\
\noalign{\vspace{3pt}}
&=&\int dx |\psi (x)|^{2}
e^{-\sum_{n}Y(a_{n},b_{n})}
e^{-\int dt\beta(t) \langle\triangle H_D(t)\rangle}.
\label{eq:cal2}
\end{eqnarray}

Consider now the weak-coupling, macroscopic limit (\ref{eq:rimba}), 
together with (\ref{eq:regg-}).
Obviously $a_{n},b_{n}\sim O(1/\sqrt{N})\longrightarrow0$ and keeping only 
terms up to order $1/N$ in $Y(a_{n},b_{n})$, we obtain 
\andy{limy}
\begin{eqnarray}
&&Y(a_{n},b_{n})\nonumber\\
&\longrightarrow &\!
a_{n}\left(1 + \frac{b_{n}^{2}}{2}\right)\nonumber\\
& &+\ln\left[
\frac{2\left(1 + b_n^2/2\right)}
{\left(1 + b_{n} + b_n^2/2\right)
+\left(1 - b_{n} + b_n^2/2\right)
\left(1 + 2a_{n}\left(1 + b_n^2/2\right) + 
(2a_n)^2/2!\right)}\right] 
\nonumber\\
\! &\simeq&\! -\frac{a_{n}^{2}}{2} + a_{n}b_{n}.
\label{eq:limy}
\end{eqnarray}
Notice that in this limit, the above quantities are expressed as 
\andy{limab,liman}
\barr
& &a_{n}^{2}\longrightarrow
\left[\hbar\omega\int dt\beta(t)\alpha_{n}(x,t)\right]
\left[\hbar\omega\int dt'\beta(t')\alpha_{n}(x,t')\right],\\
\label{eq:liman}
& & a_{n}b_{n}\longrightarrow -\hbar\omega\int dt\beta(t)\alpha_{n}^{2}(x,t).
\label{eq:limab}
\earr
Putting these results together and neglecting all border effects, we finally 
arrive at the explicit expression of the charateristic functional
\andy{cal33}
\begin{eqnarray}
\phi[\beta]
&\longrightarrow&
\int dx |\psi (x)|^{2}
e^{\frac{(\hbar\omega)^{2}}{2}\sum_{n}\int dtdt'\beta(t)\alpha_{n}(x,t)
\alpha_{n}(x,t')\beta(t')}
e^{\hbar\omega\sum_{n}\int dt\beta(t)\alpha_{n}^{2}(x,t)}\nonumber\\
& &\times
e^{-\int dt\beta(t)\hbar\omega\frac{\bar{n}}{L}(ct-x_{1})}
\nonumber\\
&\longrightarrow&
e^{\frac{1}{2}(\hbar\omega)^{2}\frac{\bar{n}}{L}
\int dtdt'\beta(t)\left[c\min(t,t')-x_{1}\right]\beta(t')}
e^{\int dt\beta(t)\hbar\omega\frac{\bar{n}}{L}(ct-x_{1})}
e^{-\int dt\beta(t)\hbar\omega\frac{\bar{n}}{L}(ct-x_{1})}
\nonumber\\
\noalign{\vspace{5pt}}
&=&e^{\frac{1}{2}(\hbar\omega)^{2}\frac{c\bar{n}}{L}
\int dtdt'\beta(t)\min(\tau,\tau')\beta(t')},
\label{eq:cal33}
\end{eqnarray}
where the interaction time $\tau=t-x_1/c$ has been introduced as before.
The characteristic functional turns out to be Gaussian, which 
proves that the stochastic process under consideration is Gaussian.
We understand that, from the appearance of $\min(\tau,\tau')$ in the 
exponent, which represents the variance of a Gaussian process, this 
process is nothing but a Wiener process. 
We stress again that this conclusion is only valid in the restricted
state-space spanned by $|\psi,0\rangle_{N}$, in the weak-coupling, macroscopic 
limit.

\setcounter{equation}{0}
\section{Conclusions and outlook } 
\label{sec-conout}
\andy{conout}

We have analyzed the modified Coleman-Hepp model and brought to light
a Wiener process in a restricted portion of the total Hilbert space.
The operator $\hh{\Sigma}(t)$ becomes a sort of ``noise operator" in the
$N\to\infty$ limit (\ref{eq:rimba}), in the sense of Eqs.\ (\ref{eq:siggmed})
and (\ref{eq:siggcorr}). 
We also proved the Gaussian white noise properties by starting from 
the characteristic functional (\ref{eq:funcar}).

Although the appearance of a stochastic process of some sort could probably be
expected on the basis of the stochastic behavior of the propagator
(\ref{eq:reess}), the emergence of the Gaussian white noise is remarkable, for
such a nontrivial Hamiltonian like (\ref{eq:H}). 
We stressed in \cite{BNNP} that the exponential decay form of the propagator
(\ref{eq:reess}) is {\em independent} of $\omega$ and depends only on $q$, the
probability of spin-flipping. 
On the contrary, the presence of $\omega$ and therefore the presence of an
energy-exchange process is essential for the derivation of the Wiener process,
through which the energy of the system is dissipated.
We understand that the weak-coupling, macroscopic limit 
(\ref{eq:rimba})-(\ref{eq:regg-}), that is closely related to van 
Hove's limit, plays a crucial role in this respect:
It corresponds to a kind of coarse graining and scale-change 
procedures, some details of which will be discussed in the Appendix.

It is useful, in this context, to briefly comment on a remark by Leggett
\cite{Leggett} that summarizes a widespread opinion among physicists 
working on these topics.
By discussing the role of the environment in connection with the collapse of 
the wave function, Leggett 
stressed the central relevance of the problem of 
dissipation to the quantum measurement theory, and argued that
``it is only genuinely dissipative processes,
in which the interaction leads to an irreversible exchange of energy between 
system and environment, which can guarantee that interference is gone beyond 
the possibility of recovery. Thus we see that it is not interaction with the 
environment as such, but specifically dissipation, which is responsible for 
genuine `decoherence': hence the central relevance of the problem of 
dissipation to quantum measurement theory".
We believe that our analysis contributes to clarify and sharpen 
the above remark: The behavior just derived, yielding a Wiener process, 
is certainly related to dephasing (``decoherence") effects 
of the same kind of those encountered in quantum measurements.
The exchange of energy between the particle and the ``environment"
(our spin system) can be considered practically irreversible.
However, the role played by $\omega$ is much more subtle, because
$\omega$ directly contributes to {\em construct} the stochastic process, as 
could be seen in Sec.\ 4. 
These remarkable features are manifest in the model here presented
and could not be guessed, in our opinion, without an explicit solution.
Another important fact, not to be dismissed, is that we have exclusively 
considered the dynamics within a restricted state space spanned by 
$\vert\psi,0\rangle_N$.
We emphasize, once again, that the stochastic process is derived
{\em without} any hypothesis on the spectrum of $H_D$ and {\em without} 
tracing over the states of the macroscopic system $D$.
This is to be contrasted to other work.

We would also like to stress that the link 
between a dissipative dynamics and quantum measurements 
is not obvious: In general, a thermal
irreversible process is a probabilistic one, described by master equations,
that characterize the approach to thermal equilibrium.
On the other hand, in a quantum measurement process, the evolution leads 
to the so-called collapse of the wave function. The final density matrix,
that does not contain off-diagonal terms, depends on the
measured observable, on the way one performs the spectral decomposition
and on the very measuring apparatus \cite{MN,Zurenv}.
The description of the loss of quantum mechanical coherence
in terms of dissipative equations, governing the evolution toward an
equilibrium situation of some sort, is therefore a delicate problem, that 
deserves further investigation.

There are other interesting open 
problems.
{}For instance, it is well known 
that the reduced dynamics of a (sub)system in interaction with a larger
system (playing the role of reservoir) is well described in terms of
quantum dynamical semigroups \cite{semigr}, so that one should be able to
derive a master (or Langevin) equation \cite{Accardi} for some dynamical
variables of the subsystem (such as the energy-momentum of the $Q$ particle). 
The derivation of a master or a Langevin equation in the present model
would open a door to thoroughly investigate a possible link between a
quantm measurement and a genuine disspative process.

 

\renewcommand{\thesection}{\Alph{section}}
\setcounter{section}{1}
\setcounter{equation}{0}
\section*{Appendix}
\label{sec-appA}
\andy{appA}

In this appendix we shall consider the more realistic situation in 
which the potential $V(x)$ has a finite width.
We shall consider, for simplicity, a square wave packet and potential
(actually, the requirement of compact support for $\psi$ and $V$ would 
suffice)
\andy{fiwi1,fiwi2}
\barr
\psi (x)&=&\frac{1}{\sqrt{2a}}
\theta(a-x)\theta(x+a)e^{ip_0x/\hbar},
\label{eq:fiwi1}\\
V(x)&=&V_0
\theta\left(\frac{\Omega}{2}-x\right)\theta\left(x+\frac{\Omega}{2}\right).
\label{eq:fiwi2}
\earr
In this case the expression (\ref{eq:alde}) for $\alpha_n$ becomes
\andy{newa}
\barr
\alpha_n&\equiv&\alpha_n(x,t)=
\frac{V_0\Omega}{\hbar c}\int_{x-x_n}^{x+ct-x_n}\frac{dy}{\Omega}
\theta\left(\frac{\Omega}{2}-y\right)\theta\left(y+\frac{\Omega}{2}\right)
\nonumber\\
&=&\frac{\sqrt{q}}{\Omega}
\left[\mbox{min}\left(\frac{\Omega}{2},x+ct-x_n\right)
-\mbox{max}\left(x-x_n,-\frac{\Omega}{2}\right)\right].
\label{eq:newa}
\earr
The expectation value of the operator $\Delta H_D(t)$ in
(\ref{eq:H't}), relative to the initial state 
$|\psi,0\rangle_N$ reads
\andy{newdh}
\barr
\langle\Delta H_D(t)\rangle&=&
\hbar\omega\int dx|\psi(x)|^2\sum_n\sin^2\alpha_n(x,t)\nonumber\\
& \longrightarrow &
\hbar\omega\frac{\bar{n}}{L}\int_{-a}^{a}\frac{dx}{2a}
\int_{x_1}^{x_N}\frac{dz}{\Omega^2}
\left[\mbox{min}\left(\frac{\Omega}{2},x+ct-z\right)
-\mbox{max}\left(x-z,-\frac{\Omega}{2}\right)\right]^2,\nonumber\\
&&
\label{eq:newdh}
\earr
where, in the last step, we have considered the weak-coupling, macroscopic 
limit (\ref{eq:rimba}). We consider now the case $ct>\Omega$ and 
focus our attention on the situation in which the wave packet is 
fully inside the potential. This means that 
\andy{requ}
\beq
x_1+\frac{\Omega}{2} +a< ct <x_N-\frac{\Omega}{2} -a.
\label{eq:requ}
\eeq
By using (\ref{eq:requ}) and since $x_1\gg\Omega,a$, 
(\ref{eq:newdh}) simply becomes
\andy{finbr}
\barr
& &\hbar\omega\frac{\bar{n}}{L}\int_{-a}^{a}\frac{dx}{2a}
\left[\int_{x_1}^{x+ct-\frac{\Omega}{2}}\frac{dz}{\Omega^2}\Omega^2
+\int_{x+ct-\frac{\Omega}{2}}^{x+ct+\frac{\Omega}{2}}
\frac{dz}{\Omega^2}\left(x+ct+\frac{\Omega}{2}-z\right)^2\right]
\nonumber\\
&=&\hbar\omega\frac{\bar{n}}{L}\int_{-a}^{a}\frac{dx}{2a}
\left(x+ct-\frac{\Omega}{6}-x_1\right)
=\hbar\omega\frac{\bar{n}}{L}\left(ct-\frac{\Omega}{6}-x_1\right).
\nonumber\\
\label{eq:finbr}
\earr
Notice that there is no effect due to the wave packet width, since it is
considered entirely inside the detector, but the finite width $\Omega$ of the
potential appears in the above formula, in contrast with (\ref{eq:medH'}).
By following a procedure similar to the previous one, we can compute the 
second order correlation function of the operator $\Delta H_D(t)$ and, 
by making use 
of the definition (\ref{eq:sigg}) of $\hh{\Sigma}(t)$ we finally obtain:
\andy{finsi1,finsi2}
\barr
\langle\hh{\Sigma}(t)\rangle&=&0,
\label{finsi1}
\nonumber\\
\langle\hh{\Sigma}(t_1)\hh{\Sigma}(t_2)\rangle& \longrightarrow &
(\hbar\omega)^2\frac{\bar{n}}{L}\biggl[\theta(\Delta t-\Omega/c)(ct_1-x_1)
\nonumber\\
& &\phantom{(\hbar\omega)^2\frac{\bar{n}}{L}\biggl[}
+\theta(\Delta t)\theta(\Omega/c-\Delta t)
\{ct_2-\Omega-x_1
-h(-\Delta t,\Omega)\}
\nonumber\\
& &\phantom{(\hbar\omega)^2\frac{\bar{n}}{L}\biggl[}
+\theta(-\Delta t)\theta(\Omega/c+\Delta t)
\{ct_1-\Omega-x_1
-h(\Delta t,\Omega)\}
\nonumber\\
& &\phantom{(\hbar\omega)^2\frac{\bar{n}}{L}\biggl[}
+\theta(-\Delta t-\Omega/c)(ct_2-x_1)\biggr],
\label{eq:finsi2}
\earr
where $\Delta t \equiv t_2-t_1$ and
\andy{deff}
\beq
h(t,\Omega)=\frac{1}{6\Omega^2}(ct+\Omega)\left( (ct+\Omega)^2-6\Omega^2 
\right) .
\label{eq:deff}
\eeq
{}From the previous two equations we can observe that, in contrast with 
(\ref{eq:siggmed}) and (\ref{eq:siggcorr}), when the finite width of the 
potential is taken in account, $\hh{\Sigma}(t)$ is not a Wiener process 
anymore, unless $\Omega\longrightarrow 0$ ($\delta$-potential limit).
Incidentally, if the time scale is changed like $t=\lambda\bar{t}$ 
(where $t$ and $\bar{t}$ can be regarded as a microscopic and a macroscopic 
time respectively) and if we define $\hh{W}(\bar{t})\equiv\hh{\Sigma}(t)$, 
then 
\andy{newwi0,newwi}
\barr
\langle\hh{W}(\bar{t})\rangle&=&0 \label{eq:newwi0} \\
\langle\hh{W}(\bar{t}_1)\hh{W}(\bar{t}_2)\rangle&=&
(\hbar\omega)^2\frac{\bar{n}}{\bar{L}}
\biggl[\theta(\Delta\bar{t}-\Omega/\lambda c)(c\bar{t}_1-\bar{x}_1)
\nonumber\\
& &\phantom{(\hbar\omega)^2\frac{\bar{n}}{\bar{L}}}
+\theta(\Delta\bar{t})\theta(\Omega/\lambda c-\Delta\bar{t})
\{c\bar{t}_2-\Omega/\lambda-\bar{x}_1
-h(-\Delta\bar{t},\Omega/\lambda)\}
\nonumber\\
& &\phantom{(\hbar\omega)^2\frac{\bar{n}}{\bar{L}}}
+\theta(-\Delta\bar{t})\theta(\Delta\bar{t}+\Omega/\lambda c)
\{c\bar{t}_1-\Omega/\lambda-\bar{x}_1
-h(\Delta\bar{t},\Omega/\lambda)\}
\nonumber\\
& &\phantom{(\hbar\omega)^2\frac{\bar{n}}{\bar{L}}}
+\theta(-\Delta\bar{t}-\Omega/\lambda c)(c\bar{t}_2-\bar{x}_1)\biggr],
\label{eq:newwi}
\earr
where $\bar{x_1}=x_1/\lambda$ and $\bar{L}=L/\lambda$. 
In this case, only when $|\bar{t}_2-\bar{t}_1|\gg\Omega/\lambda c$ (or 
$\lambda\longrightarrow\infty$ with $\bar{x_1},\bar{L},\bar{t}<\infty$, which
is equivalent to a time scale transformation), we reobtain a proper Wiener
process. In other words,  the $\delta$-potential limit can be regarded as a
realization of the macroscopic time-scale transformation. 

The above considerations bring to light the close link with 
van Hove's $\lambda^2 T$ limit, as discussed in \cite{NNP1}.
This can be easily evinced by observing that $q$, in 
Eq.\ (\ref{eq:rimba}), is nothing but the square of a coupling
constant (van Hove's $\lambda$),
and that $N (\propto L)$ can be considered proportional to
the total interaction time $T$.
Notice also that the ``lattice spacing" $d$, the inverse of which 
corresponds to a density in our 1-dimensional model, can be kept finite 
in the limit. Obviously, 
in such a case, we have to express everything in terms of scaled 
variables such as $\bar{t} = t/\lambda$, $\bar{x} = x/\lambda$ 
and $\zeta\equiv a/L$, where $a$ is the size of the wave packet.


\end{document}